\documentclass{scrartcl}

\usepackage{amsmath}
\usepackage{amssymb}
\newtheorem{theorem}{Theorem}
\usepackage{tikz}
\usepackage{algorithm}
\usepackage{algpseudocode}
\usepackage{bbm}
\usepackage{subcaption}

\definecolor{dblue}{RGB}{9,61,127}		% Hex: 093d7f
\definecolor{lblue}{RGB}{54,116,195}	% Hex: 3674c3
\definecolor{dyellow}{RGB}{204,150,52}	% Hex: cc9634

\begin{document}
	
\title{The k-d tree data structure and a proof for neighborhood computation in expected logarithmic time}
\author{Martin Skrodzki}
\date{\today}

\maketitle

%%%%%%%%%%%%%%%%%%%%%%%%%%%%%%%%%%%%%%%%%%%%%%%%%%%%%

\noindent For practical applications, any neighborhood concept imposed on a finite point set $P$ is not of any use if it cannot be computed efficiently. Thus, in this paper, we give an introduction to the data structure of k-d trees, first presented by Friedman, Bentley, and Finkel in 1977, see \cite{friedman1977algorithm}. After a short introduction to the data structure (Section~\ref{sec:TheDataStructureOfK-DTrees}), we turn to the proof of efficiency by Friedman and his colleagues (Section~\ref{sec:NeighborhoodQueriesInLogartihmicTime}). The main contribution of this paper is the translation of the proof of Freedman, Bentley, and Finkel into modern terms and the elaboration of the proof. Note that the present paper is an excerpt of a PhD thesis, see~\cite{skrodzki2019neighborhood}.

%%%%%%%%%%%%%%%%%%%%%%%%%%%%%%%%%%%%%%%%%%%%%%%%%%%%%

\section{The Data Structure of k-d Trees}
\label{sec:TheDataStructureOfK-DTrees}

In this section, we will give a brief introduction to the k-d tree data structure. It was originally presented by Jon Louis Bentley in 1975, see \cite{bentley1975multidimensional}. A modern introduction to the two-dimensional case can be found in \cite[Chapter 5.2]{berg2000computational}.

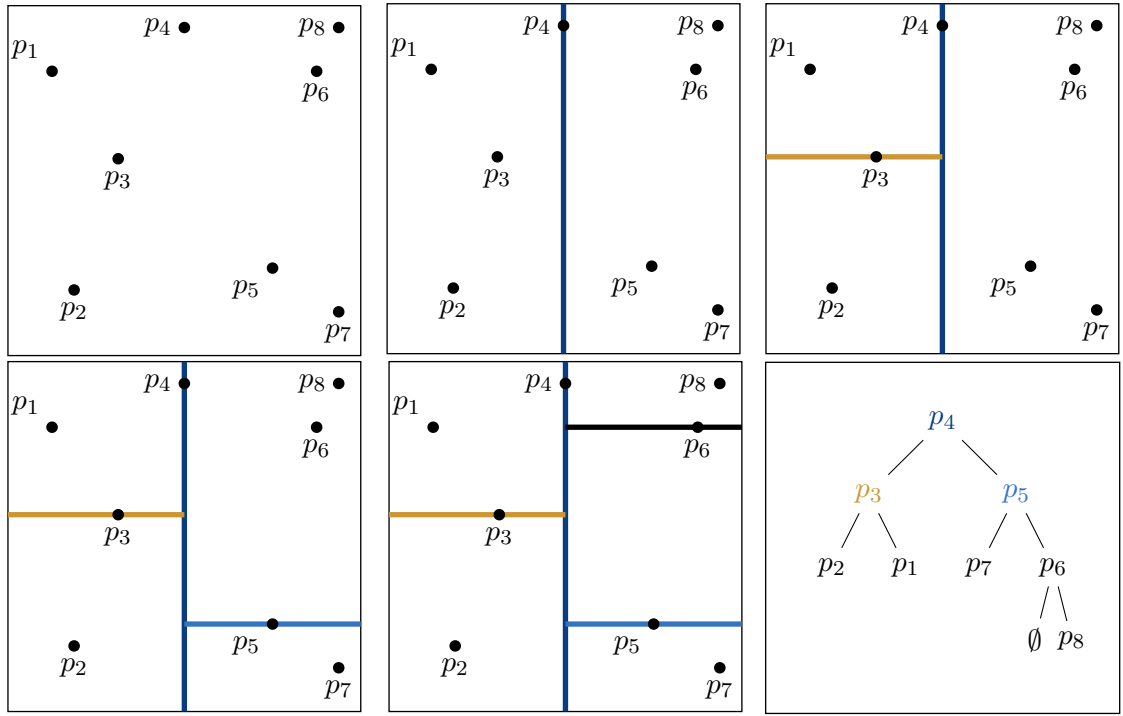
\begin{figure}
	\newcommand{\boxScale}{0.58}
	\begin{tikzpicture}[scale=\boxScale,line width = 2pt]
	\draw[line width = 0.5pt] (0,0) rectangle (8,8);
	\draw[fill=black] (1,6.5) circle (2pt) node[above left] {$p_1$};
	\draw[fill=black] (1.5,1.5) circle (2pt) node[below] {$p_2$};
	\draw[fill=black] (2.5,4.5) circle (2pt) node[below] {$p_3$};
	\draw[fill=black] (4.0,7.5) circle (2pt) node[left]  {$p_4$};
	\draw[fill=black] (6.0,2.0) circle (2pt) node[below left] {$p_5$};
	\draw[fill=black] (7.0,6.5) circle (2pt) node[below] {$p_6$};
	\draw[fill=black] (7.5,1.0) circle (2pt) node[below] {$p_7$};
	\draw[fill=black] (7.5,7.5) circle (2pt) node[left]  {$p_8$};
	\end{tikzpicture}
	\hfill
	\begin{tikzpicture}[scale=\boxScale,line width = 2pt]
	\draw[line width = 0.5pt] (0,0) rectangle (8,8);
	\draw[dblue] (4,0) -- (4,8);
	\draw[fill=black] (1,6.5) circle (2pt) node[above left] {$p_1$};
	\draw[fill=black] (1.5,1.5) circle (2pt) node[below] {$p_2$};
	\draw[fill=black] (2.5,4.5) circle (2pt) node[below] {$p_3$};
	\draw[fill=black] (4.0,7.5) circle (2pt) node[left]  {$p_4$};
	\draw[fill=black] (6.0,2.0) circle (2pt) node[below left] {$p_5$};
	\draw[fill=black] (7.0,6.5) circle (2pt) node[below] {$p_6$};
	\draw[fill=black] (7.5,1.0) circle (2pt) node[below] {$p_7$};
	\draw[fill=black] (7.5,7.5) circle (2pt) node[left]  {$p_8$};
	\end{tikzpicture}
	\hfill
	\begin{tikzpicture}[scale=\boxScale,line width = 2pt]
	\draw[line width = 0.5pt] (0,0) rectangle (8,8);
	\draw[dblue] (4,0) -- (4,8);
	\draw[dyellow] (0,4.5) -- (4,4.5);
	\draw[fill=black] (1,6.5) circle (2pt) node[above left] {$p_1$};
	\draw[fill=black] (1.5,1.5) circle (2pt) node[below] {$p_2$};
	\draw[fill=black] (2.5,4.5) circle (2pt) node[below] {$p_3$};
	\draw[fill=black] (4.0,7.5) circle (2pt) node[left]  {$p_4$};
	\draw[fill=black] (6.0,2.0) circle (2pt) node[below left] {$p_5$};
	\draw[fill=black] (7.0,6.5) circle (2pt) node[below] {$p_6$};
	\draw[fill=black] (7.5,1.0) circle (2pt) node[below] {$p_7$};
	\draw[fill=black] (7.5,7.5) circle (2pt) node[left]  {$p_8$};
	\end{tikzpicture}\\
	\begin{tikzpicture}[scale=\boxScale,line width = 2pt]
	\draw[line width = 0.5pt] (0,0) rectangle (8,8);
	\draw[dblue] (4,0) -- (4,8);
	\draw[dyellow] (0,4.5) -- (4,4.5);
	\draw[lblue] (4,2) -- (8,2);
	\draw[fill=black] (1,6.5) circle (2pt) node[above left] {$p_1$};
	\draw[fill=black] (1.5,1.5) circle (2pt) node[below] {$p_2$};
	\draw[fill=black] (2.5,4.5) circle (2pt) node[below] {$p_3$};
	\draw[fill=black] (4.0,7.5) circle (2pt) node[left]  {$p_4$};
	\draw[fill=black] (6.0,2.0) circle (2pt) node[below left] {$p_5$};
	\draw[fill=black] (7.0,6.5) circle (2pt) node[below] {$p_6$};
	\draw[fill=black] (7.5,1.0) circle (2pt) node[below] {$p_7$};
	\draw[fill=black] (7.5,7.5) circle (2pt) node[left]  {$p_8$};
	\end{tikzpicture}
	\hfill
	\begin{tikzpicture}[scale=\boxScale,line width = 2pt]
	\draw[line width = 0.5pt] (0,0) rectangle (8,8);
	\draw[dblue] (4,0) -- (4,8);
	\draw[dyellow] (0,4.5) -- (4,4.5);
	\draw[lblue] (4,2) -- (8,2);
	\draw (4,6.5) -- (8,6.5);
	\draw[fill=black] (1,6.5) circle (2pt) node[above left] {$p_1$};
	\draw[fill=black] (1.5,1.5) circle (2pt) node[below] {$p_2$};
	\draw[fill=black] (2.5,4.5) circle (2pt) node[below] {$p_3$};
	\draw[fill=black] (4.0,7.5) circle (2pt) node[left]  {$p_4$};
	\draw[fill=black] (6.0,2.0) circle (2pt) node[below left] {$p_5$};
	\draw[fill=black] (7.0,6.5) circle (2pt) node[below] {$p_6$};
	\draw[fill=black] (7.5,1.0) circle (2pt) node[below] {$p_7$};
	\draw[fill=black] (7.5,7.5) circle (2pt) node[left]  {$p_8$};
	\end{tikzpicture}
	\hfill
	\begin{tikzpicture}[scale=0.97,level distance=1.cm, level 1/.style={sibling distance=20mm},	level 2/.style={sibling distance=10mm}, level 3/.style={sibling distance=5mm}]
	\draw[line width = 0.5pt] (0,0) rectangle (4.8,4.8);
	\node[dblue] at (2.4,4) {$p_4$}
	child {node[dyellow] {$p_3$}
		child {node {$p_2$}}
		child {node {$p_1$}}	
	}
	child {node[lblue] {$p_5$}
		child {node {$p_7$}}
		child {node {$p_6$}
			child {node {$\emptyset$}}
			child {node {$p_8$}}
		}
	};
	\end{tikzpicture}
	\caption{Recursively building a k-d tree on eight points. The hyperplanes are shown in the first five figures, while the whole tree is shown in the last figure on the lower right.}
	\label{fig:kdTree2DExample}
\end{figure}

Let $P=\{p_1,\ldots,p_n\}\subset\mathbb{R}^d$ be a finite point set with $p_i=(p_i^1,\ldots,p_i^d)^T\in\mathbb{R}^d$ for ${i=1,\ldots,n}$. A k-d tree for $P$ is defined recursively. If $P$ is empty or contains only one point, an empty tree or a tree with one node containing the one point is returned. Otherwise, it is determined in which dimension $d'\in[d]$ the point set has the largest spread. That is, $d'$ is chosen such that there are two points $p_i,p_j$ with ${\left|p_i^{d'}-p_j^{d'}\right|\geq\left|p_\ell^{d''}-q_m^{d''}\right|}$ for all $d''\in[d]$, $\ell,m\in[n]$. Now, find the median\footnote{For any number $n\in\mathbb{N}$ of ordered points, we define the median to be the point with index $\lceil n/2\rceil$. Note that $\lfloor n/2\rfloor+1$ would also give a valid choice which has to be considered slightly differently when building the k-d tree.} of the points $p_i$ according to a sorting along this dimension, i.e.\ ${p_{i_1}^{d'}\leq\ldots\leq p_{i_{\lceil n/2\rceil}}^{d'}\leq\ldots\leq p_{i_n}^{d'}}$, denote it by $q=p_{i_{\lceil n/2\rceil}}$. Note that the points do not necessarily have to be sorted, as the median can be found in linear time, see \cite{blum1973time}. However, there has to be a unique ordering on the points to determine the resulting k-d tree uniquely. Thus, we assume that the point set $P$ can be uniquely ordered\footnote{\label{foot:SortingEqualPoints}In any practical application, this can be achieve by e.g.\ sorting points with equal entries according to their indices, i.e.\ in the case $p_i=p_j$ we order by $i<j$.} along any dimension $d'\in [d]$. After finding the median, a hyperplane ${H=\{x\in\mathbb{R}^d\mid x^{d'}=q^{d'}\}}$ is introduced, which splits the set $P$ into two subsets
\begin{align*}
	{P_1=\{p_{i_1},\ldots,p_{i_{\lceil n/2\rceil-1}}\}}, &&{P_2=\{p_{i_{\lceil n/2\rceil+1}},\ldots,p_{i_n}\}}
\end{align*} 
with $P_1$ containing at most one point more than $P_2$. A node is created, holding $q$ and $H$. Partitioning $P$ into the two subsets $P_1$ and $P_2$ can be performed in $\Theta(n)$. The node is given the results of recursively processing $P_1$ and $P_2$ as children and then it is returned. An example for the building process in the two-dimensional case is given in Figure~\ref{fig:kdTree2DExample}. The algorithm is given in pseudo code as Algorithm~\ref{alg:buildk-dtree}. From this building procedure, we see that the building time $T(n)$ of a k-d tree satisfies the following recursion
\begin{align*}
	T(n)=\begin{cases}
	\Theta(1) & n=1\\
	\Theta(n) + 2\cdot T(\lceil n/2\rceil) & n>1
	\end{cases},
\end{align*}
which solves to $T(n)=\Theta(n\cdot\log(n))$, see \cite[pp. 272--274]{sedgewick2011algorithms}. By the above and by noting that each node of the k-d tree stores a distinct point of the input set $P$, we proved the following theorem.

\begin{theorem}[Storage requirement and building time of a k-d tree, \cite{berg2000computational}]
	A k-d tree for a set of $n$ points uses $\Theta(n)$ storage and can be constructed in $\Theta(n\cdot\log(n))$ time.
\end{theorem}

The k-d tree data structure has applications in orthogonal range searches, as discussed in \cite[Chapter 5.2]{berg2000computational} and \cite[Chapter 3.2]{skrodzki2014neighborhood}. In the following, we will focus on applications in the context of neighborhood queries.

\begin{algorithm}
	\caption{Build k-d tree}\label{alg:buildk-dtree}
	\begin{algorithmic}[1]
		\Procedure{Build k-d tree}{point set $P$}
			\If{$|P|\leq 1$}
				\State \textbf{return} node containing $P$
			\EndIf
			\State $d'\gets$ most spread dimension of $P$
			\State $q\gets$ median according to dimension $d'$
			\State $P_1\gets\{p_i\in P\mid p_i^{d'}\leq q^{d'}, p_i\neq q\}$
			\State $P_2\gets P\backslash(P_1\cup \{q\})$
			\State $H\gets \{x\in\mathbb{R}^d\mid x^{d'}=q^{d'}\}$
			\State $N_\ell\gets$\Call{Build k-d tree}{$P_1$}
			\State $N_r\gets$\Call{Build k-d tree}{$P_2$}
			\State \textbf{return} node containing $q$, $H$ with $N_\ell$ and $N_r$ as children
		\EndProcedure
	\end{algorithmic}
\end{algorithm}

%%%%%%%%%%%%%%%%%%%%%%%%%%%%%%%%%%%%%%%%%%%%%%%%%%%%%

\section{Neighborhood Queries in Logarithmic Time}
\label{sec:NeighborhoodQueriesInLogartihmicTime}

Amongst the data structures computing the $k$~nearest neighborhood, the most prominent choice is the k-d tree data structure presented above. The reason is that in 1977, Friedmann, Bentley, and Finkel were able to prove an average case running time of $\mathcal{O}(\log(n))$ for a single neighbor query in a k-d tree built on $n$ points. The proof in their paper is very concise and covers roughly three pages \cite[pp. 214--216]{friedman1977algorithm}. Thus, we will give a more elaborate version of their proof in modern wording here.

Let a point set $P$ and a corresponding k-d tree built according to Section~\ref{sec:TheDataStructureOfK-DTrees} be given. A neighborhood query for any point $p\in\mathbb{R}^d$ is then performed by traversing the tree to the leaf representing the box which contains the query point. From there, the query goes back to the root, investigating subtrees along the path where the splitting hyperplane is closer to $p$ than the currently found nearest neighbors. When reaching a node, all elements in it are investigated as to whether they are closer to $p$ than the current closest points found. The algorithm is fast, as it can be expected that several subtrees do not have to be investigated. See Algorithm~\ref{alg:searchNN} for a pseudo code version and see \cite[Section~5.2]{skrodzki2014neighborhood} for an illustration of the nearest neighbor search for a single nearest neighbor, i.e.\ $k=1$.

\begin{algorithm}
	\caption{Nearest Neighbor Search in k-d trees}
	\label{alg:searchNN}
	\begin{algorithmic}[1]
		\item[]
		\Procedure{NNk-dTree}{point $p$, k-d tree $T$, distance $\varepsilon$, number $k$}
			\State $L\gets$empty list
			\State \textbf{return} \Call{NNk-dTreeRec}{$p$,root,$\varepsilon,k,L$}
		\EndProcedure
		\item[]
		\Procedure{NNk-dTreeRec}{point $p$, k-d tree $T$, distance $\varepsilon$, number $k$, list $L$}
			\If{$T=\emptyset$}
				\State \textbf{return} $L$
			\EndIf
			\State Extract point $p_j$ from $T$.root and store it in $L$ if $\left\|p_j-p\right\|\leq \varepsilon$.
			\If{$L$ is larger than $k$}
				\State Delete the point with largest distance to $p$ from $L$.
			\EndIf 
			\If{$T$ is just a leaf}
				\State \textbf{return} $L$
			\EndIf
			\If{$T$.root.leftSubtree contains $p$}
				\State $T_1=T$.root.leftSubtree, $T_2=T$.root.rightSubtree
			\Else
				\State $T_2=T$.root.leftSubtree, $T_1=T$.root.rightSubtree
			\EndIf
			\State \Call{NNk-dTreeRec}{$p$, $T_1$, $\varepsilon,k,L$}
			\If{$|L|<k$ \textbf{and} $\left\|p-T\text{.root.hyperplane}\right\|<\varepsilon$}
				\State\Call{NNk-dTreeRec}{$p$, $T_2$, $\varepsilon,k,L$}
			\ElsIf{$\left\|L\text{.farthest}-p\right\|>\left\|p-T.\text{root.hyperplane}\right\|$ \textbf{and} $\left\|p-T.\text{root.hyperplane}\right\|\leq \varepsilon$}
				\State\Call{NNk-dTreeRec}{$p$, $T_2$, $\varepsilon,k,L$}
			\EndIf
			\State \textbf{return} $L$
		\EndProcedure
		\item[]
	\end{algorithmic}
\end{algorithm}

We will now investigate the following question: What is the expected query time for a neighborhood query in a k-d tree? We will do this by considering a different setup compared to the construction in Section~\ref{sec:TheDataStructureOfK-DTrees}. Namely, we will store points only in the leaves of the tree and we will further allow for more than one point to be stored in each leaf. Collecting all factors that can affect the runtime of the neighborhood search, we get the following list:
\begin{itemize}
	\item total number $n$ of points,
	\item dimension $d$ of the ambient space,
	\item number of neighbors sought $k$,
	\item number $b$ of points to be stored in each leaf,
	\item distance measure\footnote{\label{foot:DistanceMeasure}While the theory holds for any distance measure, i.e. any metric, in the following we will use the Euclidean metric.} $d:\mathbb{R}^d\times\mathbb{R}^d\rightarrow\mathbb{R}_{\geq0}$, $(p,q)\mapsto d(p,q)$,
	\item density $\delta:\mathbb{R}^d\rightarrow[0,1]$ of points in space.
\end{itemize}
The density is taken into account as a way to analyze an arbitrary point set. As we will not make any assumption on the actual positions of the points $p_i\in P$ throughout the proof, we turn to a probabilistic argument. Thus, we consider an arbitrary non-empty sample space $\Omega$ and random variables $X_1,\ldots,X_n,X_p:\Omega\rightarrow\mathbb{R}^d$. Draw a sample $\omega\in\Omega$. Now, our point set $P$ is given by $P=\{X_1(\omega),\ldots,X_n(\omega)\}$ and the point $p$ to search neighbors for is given as $X_p(\omega)$. That is, we want to find $k$ points from $\{X_1(\omega),\ldots,X_n(\omega)\}\subset\mathbb{R}^d$ that are the $k$ nearest neighbors to $X_p(\omega)\in\mathbb{R}^d$ within the set $\{X_1(\omega),\ldots,X_n(\omega)\}\subset\mathbb{R}^d$. 

Without loss of generality we assume $X_1(\omega),\ldots,X_k(\omega)$ to be the $k$ nearest neighbors with $X_k(\omega)$ being the farthest from $X_p(\omega)$. Now, denote
\begin{align*}
	B_k(X_p(\omega)) = \{x\in\mathbb{R}^d\mid d(x,X_p(\omega))\leq d(X_k(\omega),X_p(\omega))\}
\end{align*}
to be the ball around $X_p(\omega)$ containing all points in $\mathbb{R}^d$ with distance less than or equal to the distance to $X_k(\omega)$. The volume of this ball is
\begin{align*}
	v_k(X_p(\omega)):=\int_{B_k(X_p(\omega))}1\:dx.
\end{align*}
Furthermore, the probability content of this region---according to the density $\delta(x)$---is
\begin{align*}
	u_k(X_p(\omega)) = \int_{B_k(X_p(\omega))}\delta(x)\:dx.
\end{align*}
Since $\int_{\mathbb{R}^d}\delta(x)\:dx=1$, we have $u_k(X_p(\Omega))\leq 1$. Furthermore, since $\delta(x)\geq 0$ for all ${x\in B_k(X_p(\omega))}$, we have ${0\leq u_k(X_p(\omega))}$. Consider some additional random variable ${X:\Omega\rightarrow\mathbb{R}^d}$ with density $\delta$. Then,
\begin{align*}
	u_k(X_p(\omega)) 	&= \int_{B_k(X_p(\omega))}\delta(x)\:dx\\
		&=\int_{\mathbb{R}^d}\delta(x)\cdot\mathbbm{1}_{B_k(X_p(\omega))}(x)\:dx\\
		&\stackrel{\diamond}{=}\mathbb{E}(\mathbbm{1}_{B_k(X_p(\omega))}(X)),
\end{align*}
where $\diamond$ holds by the law of the unconscious statistician: $\mathbb{E}(g(X))=\int_{\mathbb{R}^d}g(x)\cdot\delta(x)\:dx$, \cite[p.~156]{blitzstein2014introduction}, with $g(x):=\mathbbm{1}_{B_k(X_p(\omega))}(x)$. Furthermore, we have
\begin{align*}
	\mathbbm{1}_{B_k(X_p(\omega))}(x)=\begin{cases}1, & \text{ if }x\in B_k(X_p(\omega))\\0, & \text{ otherwise}\end{cases}.
\end{align*}
Therefore,
\begin{align*}
	\mathbb{E}(\mathbbm{1}_{B_k(X_p(\omega))}(X))
		&=\mathbb{P}(X\in B_k(X_p(\omega)))\\
		&=\mathbb{P}(d(X,X_p(\omega))\leq d(X_k(\omega),X_p(\omega))).
\end{align*}

Now, from $X$ and $X_i$ we define new random variables 
\begin{align*}
	\xi:=d(X,X_p(\omega)), && \xi_i:=d(X_i,X_p(\omega)),
\end{align*}
which provide distances
\begin{align*}
	\xi_i(\omega):=d(X_i(\omega),X_p(\omega)).
\end{align*}  
Then, ${u_k(X_p(\omega))=\mathbb{P}(\xi\leq\xi_k(\omega))}$ by the above computations. Furthermore, ${\xi:\Omega\rightarrow\mathbb{R}}$ is a random variable with distribution ${F_\xi(x)=\mathbb{P}(\xi\leq x)}$, therefore ${u_k(X_p(\omega))=F_\xi(\xi_k(\omega))}$. The question to be answered now is: What is the distribution $F_\xi$ at $\xi_k(\omega)$?

Order the distances $\xi_i(\omega)$, then we have the following one-to-one correspondence:
\begin{align*}
	\xi_1(\omega)\leq\ldots\leq\xi_n(\omega) \stackrel{1:1}{\longleftrightarrow} F_\xi(\xi_1(\omega))\leq \ldots\leq F_\xi(\xi_n(\omega)).
\end{align*}
Now, consider the random variable $F(\xi_i):\Omega\rightarrow[0,1]$. It is defined by
\begin{align*}
	F(\xi_i)(\omega):=F_\xi(\xi_i(\omega))
\end{align*} 
and its distribution is
\begin{align*}
	\mathbb{P}(F(\xi_i)\leq x)=\mathbb{P}(\xi_i\leq F_{\xi}^{-1}(x))\stackrel{\star}{=}F_{\xi_i}(F_{\xi}^{-1}(x))\stackrel{\diamond}{=}x,
\end{align*}
where $\star$ holds because $F_{\xi_i}(x)=\mathbb{P}(\xi_i\leq x)$ and where $\diamond$ holds because $\xi$ and $\xi_i$ have the same density $\delta$ and thus the same distribution $F_\xi=F_{\xi_i}$. Hence, a single $F(\xi_i)$ is uniformly distributed. However, when choosing an ordering, the $k$-th distance $\xi_k$ is $\beta$-distributed (see appendix on page~\pageref{sec:BetaDistribution}), therefore
\begin{align*}
	u_k(X_p(\omega))=\beta(\omega).
\end{align*}
It can be shown that 
\begin{align}
	\label{equ:ExpectedVolume}
	u_k(X_p(\omega))=\frac{k}{(n+1)},
\end{align}
see appendix on page~\pageref{sec:BetaDistribution}. This states that any compact volume enclosing exactly $k$ points has probability content $\frac{k}{(n+1)}$ on average. Now, assume that $n$ is large enough such that $B_k(X_p(\omega))$ is small and thus $\delta(x)$ can be approximated by a constant $\bar{p}(X_p(\omega))$ on $B_k(X_p(\omega))$. In this case
\begin{align}
	\label{equ:DensityIntegral}
	\int_{B_k(X_p(\omega))}\delta(x)\:dx = u_k(X_p(\omega))\approx\bar{p}(X_p(\omega))\cdot v_k(X_p(\omega)).
\end{align}
Assume further that $\delta(x)$ is continuous (which enables us to approximate it over a small region in the first place), then by definition of $B_k(X_p(\omega))$, there are small neighborhoods $B_{\varepsilon_i}(X_i(\omega))$, $i=1,\ldots, k-1$ such that $\delta(x)>0$ for all $x\in B_{\varepsilon_i}(X_i(\omega))$, $i=1,\ldots,k-1$. Therefore,
\begin{align*}
	\bar{p}(X_p(\omega))=\int_{B_k(X_p(\omega))}\delta(x)\:dx \stackrel{\delta(x)\geq0\:\forall x\in\mathbb{R}^d}{\geq} \sum_{i=1}^{k-1}\underbrace{\int_{B_{\varepsilon_i}(X_i(\omega))}\delta(x)\:dx}_{>0}>0.
\end{align*}
Now, we obtain
\begin{align*}
	\frac{k}{(n+1)}&\stackrel{(\ref{equ:ExpectedVolume})}{=}u_k(X_p(\omega)) \stackrel{(\ref{equ:DensityIntegral})}{\approx}\bar{p}(X_p(\omega))\cdot v_k(X_p(\omega)),
\end{align*}
which is equivalent to
\begin{align}
	\label{equ:ExpectedBallVolumes}
	v_k(X_p(\omega))\approx\frac{k}{(n+1)\cdot\bar{p}(X_p(\omega))}.
\end{align}
Observe now the effect of the k-d tree partitioning algorithm described in Section~\ref{sec:TheDataStructureOfK-DTrees}. Choosing the median ensures that the bucket sizes of all non-empty buckets will be between $\lceil b/2\rceil$ and $b$, where $b$ is the maximum bucket size. Choosing to split on the widest spread dimension ensures that the geometric shape of these buckets will be reasonably compact. In fact, the expected edge lengths of these buckets at most differ pairwise by a factor of $2$. The buckets themselves are by this approximation hypercubical with edge length equal to the $d$-th root of the volume of space occupied by the bucket. The edges are parallel to the coordinate axes. The effect of the k-d tree partitioning, then, is to divide the coordinate space into approximately hypercubical subregions, each containing the aforementioned roughly same number of records. Because of (\ref{equ:ExpectedBallVolumes}) we have that the expected volume of such a bucket is
\begin{align}
	\label{equ:ExpectedBucketVolume}
	\mathbb{E}(v_b(X_b))\approx\frac{b}{(n+1)\cdot\bar{p}(X_b)},
\end{align}
where $X_b:\Omega\rightarrow\mathbb{R}^d$ is a random variable yielding a point that locates the bucket in the coordinate space.

\begin{figure}
	\centering	
	\begin{subfigure}[t]{0.3\textwidth}
		\centering
		\begin{tikzpicture}
			\draw (0,0) rectangle (2,2);
			\draw (1,1) circle (1);
		\end{tikzpicture}
		\caption{Ball $B_k$ and Cube $C_k$ for ${d=2}$.}
		\label{fig:BallsAndCubeD2}
	\end{subfigure}
	\hfill
	\begin{subfigure}[t]{0.3\textwidth}
		\centering
		\begin{tikzpicture}
			% Draw the hypercube
			\draw (0,0) rectangle (2,2);
			\draw[dashed] (0,0) -- (0.7,0.7);
			\draw (0,2) -- (0.7,2.7);
			\draw (2,2) -- (2.7,2.7);
			\draw (2,0) -- (2.7,0.7);
			\draw[dashed] (0.7,2.7) -- (0.7,0.7) -- (2.7,0.7);
			\draw (0.7,2.7) -- (2.7,2.7) -- (2.7,0.7);
			% Draw a sphere within the hypercube
			\shade[ball color = gray!40, opacity = 0.9] (1.35,1.35) circle (1);
		\end{tikzpicture}
		\caption{Ball $B_k$ and Cube $C_k$ for ${d=3}$.}
		\label{fig:BallsAndCubeD3}
	\end{subfigure}
	\hfill
	\begin{subfigure}[t]{0.3\textwidth}
		\centering
		\begin{tikzpicture}
		\draw (0,0) rectangle (2,2);
		\draw (1,1) circle (1);
		\draw[dashed] (-0.5,-0.5) rectangle (1,1);
		\draw[dashed] (-0.5,1) rectangle (1,2.3);
		\draw[dashed] (1,1.9) rectangle (2.5, 2.5);
		\draw[dashed] (1,1.6) rectangle (2.7, 1.9);
		\draw[dashed] (1,-0.3) rectangle (2.3, 1.6);
		\end{tikzpicture}
		\caption{A hypercube $C_k$ overlapped by $\bar{\ell}=5$ buckets. }
		\label{fig:HypercubeOverlapped}
	\end{subfigure}
	\caption{Illustration of hypercubes in different dimensions.}
\end{figure}
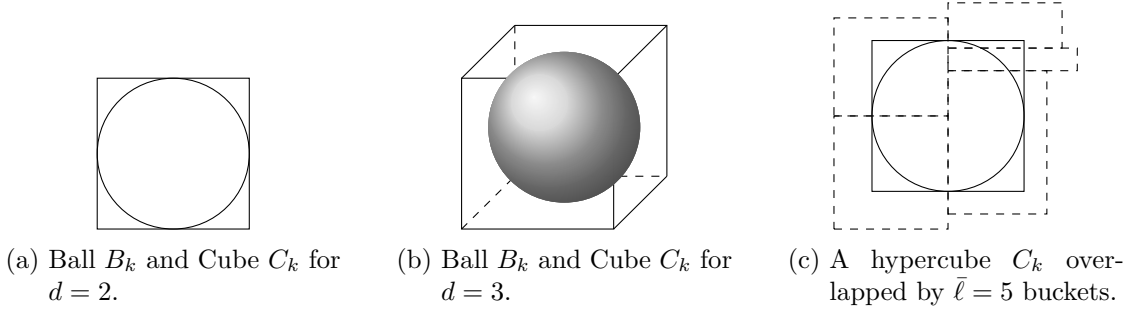

Consider now the smallest $d$-dimensional hypercube $C_k(X_p(\omega))$ with edges parallel to the coordinate axes that completely contains the ball $B_k(X_p(\omega))$. The volume $V_k(X_p(\omega))$ of this hypercube is proportional to $v_k(X_p(\omega))$, with proportionality constant
\begin{align*}
	\frac{\text{volume}(d-\text{dim cube})}{\text{volume}(d-\text{dim ball})}=\frac{(2r)^d}{r^d\cdot\frac{\pi^{d/2}}{\Gamma(d/2+1)}}=\frac{2^d\cdot\Gamma(d/2+1)}{\pi^{d/2}}=:G(d),
\end{align*}
where $r$ is the radius of the ball. See Figures~\ref{fig:BallsAndCubeD2} and~\ref{fig:BallsAndCubeD3} for an illustration of $B_k$ and $C_k$.
Therefore,
\begin{align}
	\label{equ:ExpectedHypercubeVolume}
		V_k(X_p(\omega)) &= G(d)\cdot v_k(X_p(\omega)) \stackrel{(\ref{equ:ExpectedBallVolumes})}{\approx} \frac{G(d)\cdot k}{(n+1)\cdot\bar{p}(X_p(\omega))}.
\end{align}

In order to calculate the average number of buckets $\bar{\ell}$ examined by the k-d tree during a search for the $k$ nearest neighbors, we need to find the buckets overlapping the ball $B_k(X_p(\omega))$, see Figure~\ref{fig:HypercubeOverlapped} for an illustration. This number $\bar{\ell}$ is bounded from above by $\bar{L}$, the number of buckets overlapping the hypercube $C_k(X_p(\omega))$, where 
\begin{align*}
	\bar{L} = \begin{cases}\left(\left\lfloor\frac{e_k(X_p(\omega))}{e_b(X_p(\omega))}\right\rfloor+1\right)^d & \text{ for }e_b(X_p(\omega))\leq e_k(X_p(\omega))\\ 2^d & \text{ otherwise }\end{cases}
\end{align*}
where $e_k(X_p(\omega))$ denotes the edge length of $C_k(X_p(\omega))$ and $e_b(X_p(\omega))$ denotes the edge length of the buckets in the neighborhood. However, the edge length of a hypercube is the $d$-th root of its volume, hence
\begin{align*}
	e_k(X_p(\omega))=\sqrt[d]{V_k(X_p(\omega))}, && e_b(X_p(\omega))=\sqrt[d]{V_b(X_b(\omega))}.
\end{align*}

Assume that buckets around $B_k(X_p(\omega))$ are smaller than $B_k(X_p(\omega))$ and since they are close, $\bar{p}(X_p(\omega))\approx\bar{p}(X_b(\omega))$. Then, by (\ref{equ:ExpectedHypercubeVolume}) we have
\begin{align*}
	e_k(X_p(\omega))\approx\sqrt[d]{\frac{k\cdot G(d)}{(n+1)\bar{p}(X_p(\omega))}}, && e_b(X_p(\omega))=\sqrt[d]{\frac{b\cdot G(d)}{(n+1)\bar{p}(X_b(\omega))}}.
\end{align*}
Finally,
\begin{align*}
	\bar{l}\leq\bar{L}&=\left(\left\lfloor\frac{e_k(X_p(\omega))}{e_b(X_p(\omega))}\right\rfloor+1\right)^d \approx \left(\left\lfloor\frac{\sqrt[d]{\frac{k\cdot G(d)}{(n+1)\bar{p}(X_p(\omega))}}}{\sqrt[d]{\frac{b\cdot G(d)}{(n+1)\cdot\bar{p}(X_b(\omega))}}}\right\rfloor+1\right)^d\\
	&=\left(\left\lfloor\sqrt[d]{\frac{k}{b}}\right\rfloor+1\right)^d\leq\left(\left(\frac{k}{b}\right)^{\frac{1}{d}}+1\right)^d
\end{align*}
is an upper bound for the average number of buckets overlapping $B_k(X_p(\omega))$. Here, we use that $B_k(X_p(\omega))$ is small and thus we have ${X_p(\omega)\approx X_b(\omega)}$. Note that the inequality holds as $\sqrt[d]{k/b}\geq 0$. The number of records in each bucket is $b$, so an upper bound on the number of records examined is
\begin{align*}
	\bar{R}\leq b\cdot\bar{L} \leq b\cdot\left(\left(\frac{k}{b}\right)^{1/d}+1\right)^d = \left(k^{1/d}+b^{1/d}\right)^d.
\end{align*}
This expression is minimized when choosing $b=1$ (which explains the corresponding choice in Section~\ref{sec:TheDataStructureOfK-DTrees}), yielding
\begin{align}
\label{equ:NumberOfBuckets}
	\bar{R}\leq\left(k^{1/d}+1\right)^d,
\end{align}
which is independent of the number of records $n$ and the density $\delta(x)$.

The constancy of the number of records examined as the number of records increases implies that the time required for a nearest neighbor search is equal to finding the record in the balanced binary tree, the k-d tree, which takes $ \mathcal{O}(\log(n)) $ on average. Thus, the following theorem is proven:

\begin{theorem}[Runtime of nearest neighbor queries in a k-d tree, \cite{friedman1977algorithm}]
	\label{the:RuntimeOfNearestNeighborQueryInKdTrees}
	The expected query time for a neighborhood query in a k-d tree is $\mathcal{O}(\log(n))$.
\end{theorem}

%%%%%%%%%%%%%%%%%%%%%%%%%%%%%%%%%%%%%%%%%%%%%%%%%%%%%

\section*{Conclusion}

We have presented the data structure of k-d trees and have given the classical proof of Friedman, Bentley, and Finkel in a modernized form. The proof shows the practical relevance of k-d trees for the field of geometry processing, as both combinatorial and metric neighborhood concepts can be computed efficiently utilizing the k-d tree data structure.

%%%%%%%%%%%%%%%%%%%%%%%%%%%%%%%%%%%%%%%%%%%%%%%%%%%%%

\section*{Appendix: Beta Distribution}
\label{sec:BetaDistribution}

The proof in the Section~\ref{sec:NeighborhoodQueriesInLogartihmicTime} makes use of the Beta Distribution which we will present here briefly with the facts used in the proof above. We follow the description in \cite[Section 2.5.3]{georgii2015stochastik}. It describes the waiting time for the $k$-th event of $n$ events happening in $(0,1)\subset\mathbb{R}$, called $1,\ldots, n$. Hence, the space of results as well as the sample space is $\Omega=(0,1)^n$. For all $1\leq i\leq n$, $\omega\in(\omega_1,\ldots,\omega_n)\in\Omega$, let $T_i(\omega)=\omega_i$ be the time of event $i$. Assume all $\omega_i$ are uniform distributed in $(0,1)$.

Now order the $T_i(\omega)$ strictly, which is possible since $\mathbb{P}(\bigcup_{i\neq j}\{\omega_i=\omega_j\})=0$, i.e.\ no two events happen at the same time. Call the order
\begin{align*}
T_{1:n}<\ldots<T_{n:n}\text{ with }\{T_{1:n},\ldots,T_{n:n}\}=\{T_1(\omega),\ldots,T_n(\omega)\},
\end{align*}
then $T_{k:n}$ is the time of the $k$-th event. Now fix $k$, $n$, and some $c\in(0,1)$. For a given order $T_{1:n}<\ldots<T_{n:n}$, we obtain
\begin{align*}
\underbrace{\mathbb{P}(T_{k:n}\leq c)}_{\text{fixed order}} &=\int_0^1\ldots\int_0^1 \mathbbm{1}_{\{t_1<\ldots<t_n\}}(t_1,\ldots,t_n)\cdot\mathbbm{1}_{(0,c]}(t_k)\:dt_n\ldots dt_1.
\end{align*}
For any other order, we can exchange the integration by Fubini, s.t.\ it yields the same expression independent of any of the $n!$ possible orders. Hence,
\begin{align}
\label{equ:ProbabilityOfKthEvent}
\begin{split}
\mathbb{P}(T_{k:n}\leq c) &=n!\int_0^1\ldots\int_0^1 \mathbbm{1}_{\{t_1<\ldots<t_n\}}(t_1,\ldots,t_n)\cdot\mathbbm{1}_{(0,c]}(t_k)\:dt_n\ldots dt_1\\
&=n!\cdot\underbrace{\int_0^1\mathbbm{1}_{(0,c]}(t_k)}_{=\int_0^c 1}\\
&\phantom{=}\cdot \left[\underbrace{\left(\int_0^1\ldots\int_0^1\mathbbm{1}_{\{t_1<\ldots<t_k\}}(t_1,\ldots,t_k)\:dt_{k-1}\ldots dt_1\right)}_{\int_0^{t_k}\ldots\int_0^{t_k}\mathbbm{1}_{\{t_1<\ldots<t_k\}}(t_1,\ldots,t_{k-1})\:dt_{k-1}\ldots dt_1}\right.\\
&\phantom{=}\cdot\left.\underbrace{\left(\int_0^1\ldots\int_0^1\mathbbm{1}_{\{t_k<\ldots<t_n\}}(t_k,\ldots,t_n)\:dt_{k+1}\ldots dt_n\right)}_{=\int_{t_k}^1\ldots\int_{t_k}^1\mathbbm{1}_{\{t_{k+1}<\ldots<t_n\}}(t_{k+1},\ldots,t_n)\:dt_n\ldots dt_{k+1}}\right]
\end{split}
\end{align}
Using the following two identities, see \cite[p.~44]{georgii2015stochastik},
\begin{align*}
\int_0^s\ldots\int_0^s\mathbbm{1}_{\{t_1<\ldots<t_{k-1}\}}(t_1,\ldots,t_{k-1})\:dt_{k-1}\ldots dt_1 &=\frac{s^{k-1}}{(k-1)!}\\
\int_0^s\ldots\int_0^s\mathbbm{1}_{\{t_{k+1}<\ldots<t_{n}\}}(t_{k+1},\ldots,t_{n})\:dt_{n}\ldots dt_{k+1} &=\frac{(1-s)^{n-k}}{(n-k)!}
\end{align*}
in (\ref{equ:ProbabilityOfKthEvent}), we get
\begin{align*}
\mathbb{P}(T_{k:n}\leq c)=\frac{n!}{(k-1)!(n-k)!}\cdot\int_0^cs^{k-1}(1-s)^{n-k}\:ds.
\end{align*}
In general, for a random variable $X$ with distribution $F_X$ and density $f$, if $F_X$ is differentiable, we have $f(x)=F'(X)$. Finally, it can be shown, see \cite[p.~45]{georgii2015stochastik}, that $\mathbb{E}(\beta_{a,b})=\frac{a}{a+b}$, where for $k,n\in\mathbb{N}$, $\beta_{k,n-k+1}$ describes the distribution of the $k$-th smallest time of $n$ random times in $[0,1]$. In particular, $\beta_{1,n}(s)=n(1-s)^{n-1}$ is the density of the first time and $\beta_{1,1}=\mathcal{U}_{(0,1)}$.

%%%%%%%%%%%%%%%%%%%%%%%%%%%%%%%%%%%%%%%%%%%%%%%%%%%%%

\end{document}